\documentclass[12pt]{article}

\usepackage{amsmath}
\usepackage{cite}
\usepackage[xy,curve,matrix,oldcm,debug]{qsymbols}
\usepackage{hhline}

\numberwithin{equation}{section}

\title{Lagrangian formulation of free arbitrary $N$-extended massless higher spin
supermultiplets in 4D, AdS space}

\author{I.L. Buchbinder${}^{ab}$\thanks{joseph@tspu.edu.ru},
T.V. Snegirev${}^{a}$\thanks{snegirev@tspu.edu.ru}
\\[0.5cm]
\it{\small ${}^a$Department of Theoretical Physics, Tomsk State
Pedagogical University,}\\
\it{\small Tomsk, 634061, Russia}\\
\it{\small ${}^b$National Research Tomsk State University, Tomsk
634050, Russia}}

\date{}

\begin{document}

\maketitle

\begin{abstract}
We derive the component Lagrangian for the free $N$-extended
on-shell massless higher spin supermultiplets in four-dimensional
anti-de Sitter space. The construction is based on frame-like
description of massless integer and half-integer higher spin fields.
The massless supermultiplets are formulated for $N\leq 4k,$ where
$k$ is a maximal integer or half-integer spin in the multiplet. The
supertransformations that leave the Lagrangian invariant are found
in explicit form and it is shown that their algebra is closed
on-shell.

\end{abstract}

\thispagestyle{empty}
\newpage
\setcounter{page}{1}

\section{Introduction}

The study of various aspects of higher spin fields is currently one
of the actively developing areas of modern theoretical and
mathematical physics (for review see e.g.
\cite{Vas04,FT09,BBS10,DS14,Vas15}). Recently, there has been a
surge of interest in constructing the supersymmetric higher spin
models\footnote{Higher spin models are sometimes called in the
literature the higher spin (super)gravities.} and investigating the
properties of such models
\cite{BK15,Kuz16a,Kuz16b,Kuz17a,Kuz17b,Kuz17c,Kuz18a,Kuz18b,Kuz18c,Kuz18d,Kuz18e,Kuz18f,
Kuz19a,Kuz19b,Kuz19c,Kuz19d,Kuz20a,Kuz20b,BGKa,BGKb,BGKc,BGKd,BGKe,Koutra,Koutrb,
Koutrc,Koutrd,ST,BSZ-3a,BSZ-3b,BSZ-3c,BSZ-3d,BSZa,BSZb,BSZc,KhZ,Met1,Met2}.
In this paper, we study a general problem of Lagrangian construction
for arbitrary $N$-extended massless free on-shell supermultiplets in
four dimensional $AdS$ space and derive the Lagrangians describing
the dynamics of such supermultiplets.

It is well known that in four dimensions the $N$-extended
supermultiplets with maximal spin $k=1$ are restricted by the
condition $N \leq 4$ and the multiplets with maximal spin $k=2$ are
restricted by the condition $N \leq 8$. The supermultiplets with
$N>8$ must contain the higher spins $k>2$. To be more precise, there
is a specific relationship between the parameter $N$ and the highest
spin $k$ in the supermultiplet, $N \leq4k$ (see e.g. \cite{Sohn}).
Of course, if one does not restrict the maximal spin in the
multiplet by the quantities $k = 1,2,$ then for any N there exist
the supermultiplets with arbitrary higher spins.

For the case of simple $N=1$ supersymmetry,  the component
Lagrangian formulation of on-shell higher spin supermultiplets in
Minkowski space has been known for a long time \cite{Cur79,Vas80}
and further the component approach has been generalized and studied
in the works \cite{Vas87,
KonVas90,Engquist:2002vr,Sezgin:1998gg,Sezgin:12SS,Alkalaev:2002rq,Zin}.
In particular, supertransformations were found that leave invariant
the sum of Lagrangians for free massless fields with spins $k$ and
$k+1/2$. Completely off-shell Lagrangian formulation for such
theories was constructed within the framework of the superfield
approach \cite{KSP93,KS93}\footnote{See also the later paper
\cite{GK} on the same subject.}. Off-shell formulation of the N=1
higher superspin free Lagrangian theory in 4D, AdS space has been
first developed in the work \cite{KS94} from the very beginning in
superfield formalism and its component form was derived from
superfield theory. Quantization of this theory has been given in
\cite{BKS}. The $N=2$ supersymmetric higher spin models both in the
Minkowski and AdS spaces were discussed in \cite{GKS96a}, the
universal higher spin superfield approach in 4D, $N=1$ AdS
superspace has been developed in \cite{GKS97}.

Recently the on-shell superfield Lagrangian realization was
constructed for the extended $N=1$ massless supermultiplets in the
framework of light-cone gauge formalism \cite{Met1}. Extension of
this approach for on-shell $N$-extended superfield Lagrangian
formulation has been given in \cite{Met2} under the condition
$N=4n$, where $n$ is a natural number. In this paper we generalize
the results of the work \cite{Met2} and give an explicit component
Lagrangian construction of arbitrary $N$-extended massless higher
spin on-shell supermultiplets in the four-dimensional anti-de Sitter
space without using the condition accepted in \cite{Met2}.

Our construction is based on the frame-like approach for higher spin
fields. The generic scheme of the Lagrangian formulation for free
higher spin bosonic and fermionic fields in this approach was
developed in \cite{Vas87}. The full higher spin field Lagrangian is
sum of the free Lagrangians for the bosonic and fermionic component
fields of the on-shell supermultiplet. The main question that must
be solved in such an approach is finding the supersymmetry
transformations that leave the full Lagrangian invariant. In
principle, the necessary supersymmetry transformations can be
obtained on the base of the construction developed in
\cite{KonVas90,Sezgin:12SS}, however an explicit realization of the
supersymmetry transformations is not derived so far. In the given
paper we fill this gap and find the explicit supertransformations
for $N$-extended massless higher spin supermultiplets that leave the
sum of free bosonic and fermionic Lagrangians invariant and show
that the algebra of the supertransformations is closed on-shell.

The paper is organized as follows. In Section \ref{Section1} we
describe the basic elements of the frame-like Lagrangian formulation
for free massless higher spin fields in $4D$ AdS space and the $4D$
multispinor technique. In Section \ref{Section2} we present the
minimal massless $N=1$ supermultiplets  \cite{BSZa} which will be
used as the building blocks to construct the $N$-extended
supermultiplets. Section \ref{Section3} is devoted to constructing
the arbitrary $N$-extended massless supermultiplets in $4D$ AdS. For
each case, we formulate the field contents and introduce the
corresponding field variables. Then we derive the
supertransformations for these supermultiplets and define the
Lagrangian as a sum of Lagrangians for all integer and half-integer
spin fields of the given supermultiplet. We prove that such a
Lagrangian is invariant under the above transformations. Finally, we
show that the constructed supertransformations form the closed
$N$-extended $4D$ AdS superalgebras.

\section{Free higher spin fields}\label{Section1}
In this section, we briefly consider the frame-like Lagrangian
formulation of the free massless higher spin fields in the $4D$ AdS
space and the corresponding $4D$ multispinor formalism.

In the frame-like approach, the massless fields with integer spin
$k\geq2$ are described by the dynamical 1-form
$f^{\alpha(k-1)\dot\alpha(k-1)}$ and the auxiliary 1-form
$\Omega^{\alpha(k)\dot\alpha(k-2)}$,
$\Omega^{\alpha(k-2)\dot\alpha(k)}$ (see all notations in the
appendix). These fields are totally symmetric with respect to the
dotted and undotted indices and generalize the tetrad field and
Lorentz connection in the frame formulation of gravity. We choose
them to bee real valued that is they satisfy the following rules of
hermitian conjugation
\begin{eqnarray*}
(f^{\alpha(k-1)\dot\alpha(k-1)})^\dag&=&f^{\alpha(k-1)\dot\alpha(k-1)},
\\
(\Omega^{\alpha(k)\dot\alpha(k-2)})^\dag&=&\Omega^{\alpha(k-2)\dot\alpha(k)}.
\end{eqnarray*}
The Lagrangian being the differential 4-form in $4D$  AdS space
looks like
\begin{eqnarray}\label{BosLag}
\frac{(-1)^k}{i}{\cal
L}_{k}&=&k\Omega^{\alpha(k-1)\beta\dot\alpha(k-2)}E_\beta{}^\gamma
\Omega_{\alpha(k-1)\gamma\dot\alpha(k-2)}
-(k-2)\Omega^{\alpha(k)\dot\alpha(k-3)\dot\beta}E_{\dot\beta}{}^{\dot\gamma}
\Omega_{\alpha(k)\dot\alpha(k-3)\dot\gamma}\nonumber\\
&&+2\Omega^{\alpha(k-1)\beta\dot\alpha(k-2)}e_\beta{}^{\dot\beta}
Df_{\alpha(k-1)\dot\alpha(k-2)\dot\beta}\nonumber\\
&&+ 2k\lambda^2f^{\alpha(k-2)\beta\dot\alpha(k-1)}E_\beta{}^\gamma
f_{\alpha(k-2)\gamma\dot\alpha(k-1)}+h.c. \,\,.
\end{eqnarray}
Here 1-form $e^{\alpha\dot\alpha}$ is the AdS background tetrad, $D$
is the AdS covariant derivative $De^{\alpha\dot\alpha}=0$,
$E^{\alpha\beta}$ and $E^{\dot\alpha\dot\beta}$ are a double product
of $e^{\alpha\dot\alpha}$ (see appendix for details). The form of
Lagrangian (\ref{BosLag}) is determined by the invariance under the
gauge transformations
\begin{eqnarray*}
\delta f^{\alpha(k-1)\dot\alpha(k-1)}&=&
D\xi^{\alpha(k-1)\dot\alpha(k-1)}
+e_\beta{}^{\dot\alpha}\eta^{\alpha(k-1)\beta\dot\alpha(k-2)}+
e^\alpha{}_{\dot\beta}\eta^{\alpha(k-2)\dot\alpha(k-1)\dot\beta},
\\
\delta
\Omega^{\alpha(k),\dot\alpha(k-2)}&=&D\eta^{\alpha(k),\dot\alpha(k-2)}
+\lambda^2
e^\alpha{}_{\dot\beta}\xi^{\alpha(k-1)\dot\alpha(k-2)\dot\beta},
\\
\delta
\Omega^{\alpha(k-2),\dot\alpha(k)}&=&D\eta^{\alpha(k-2),\dot\alpha(k)}
+\lambda^2
e_\beta{}^{\dot\alpha}\xi^{\alpha(k-2)\beta\dot\alpha(k-1)}.
\end{eqnarray*}
The remarkable property of frame-like formulation is possibility to
construct the gauge invariant objects which generalize the torsion
and curvature in gravity
\begin{eqnarray*}
{\cal
T}^{\alpha(k-1)\dot\alpha(k-1)}&=&Df^{\alpha(k-1)\dot\alpha(k-1)}
+e_\beta{}^{\dot\alpha}\Omega^{\alpha(k-1)\beta\dot\alpha(k-2)}+
e^\alpha{}_{\dot\beta}\Omega^{\alpha(k-2)\dot\alpha(k-1)\dot\beta},
\\
{\cal
R}^{\alpha(k),\dot\alpha(k-2)}&=&D\Omega^{\alpha(k),\dot\alpha(k-2)}
+\lambda^2
e^\alpha{}_{\dot\beta}f^{\alpha(k-1)\dot\alpha(k-2)\dot\beta},
\\
{\cal
R}^{\alpha(k-2),\dot\alpha(k)}&=&D\Omega^{\alpha(k-2),\dot\alpha(k)}
+\lambda^2
e_\beta{}^{\dot\alpha}f^{\alpha(k-2)\beta\dot\alpha(k-1)}.
\end{eqnarray*}
To simplify the construction of the supermultiplets we do not
introduce any supertransformations for the auxiliary fields
$\Omega$. Instead, all calculations are done up to the terms
proportional to the auxiliary field equations of motion. It is
equivalent to the following "zero torsion conditions":
\begin{equation}\label{TorsHS}
{\cal T}^{a(k-1)}\approx0\quad\Rightarrow\quad
e_\beta{}^{\dot\alpha}{\cal R}^{\alpha(k-1)\beta\dot\alpha(k-2)}+
e^\alpha{}_{\dot\beta}{\cal
R}^{\alpha(k-2)\dot\alpha(k-1)\dot\beta}\approx0 \, .
\end{equation}
As for the supertransformations for the dynamical fields $f$, the
corresponding variation of the Lagrangian can be compactly written
as follows
\begin{eqnarray*}
(-1)^k\delta{\cal L}_{k}&=&-i2{\cal
R}^{\alpha(k-1)\beta\dot\alpha(k-2)}e_\beta{}^{\dot\beta} \delta
f_{\alpha(k-1)\dot\alpha(k-2)\dot\beta}+h.c.
\end{eqnarray*}

Now let us turn to massless fields with half-integer spin
$k+1/2\geq3/2$ which are described by 1-form
$\Phi^{\alpha(k)\dot\alpha(k-1)}$,
$\Phi^{\alpha(k-1)\dot\alpha(k)}$. To be Majorana fields they must
satisfy the reality condition
$$
(\Phi^{\alpha(k)\dot\alpha(k-1)})^\dag=\Phi^{\alpha(k-1)\dot\alpha(k)}.
$$
The corresponding Lagrangian has form
\begin{eqnarray}\label{FerLag}
(-1)^k{\cal
L}_{k+\frac12}&=&\Phi_{\alpha(k-1)\beta\dot\alpha(k-1)}e^\beta{}_{\dot\beta}
D\Phi^{\alpha(k-1)\dot\alpha(k-1)\dot\beta}\nonumber\\
&&+\epsilon_{k+\frac12}\frac{\lambda}{2}[(k+1)
\Phi_{\alpha(k-1)\beta\dot\alpha(k-1)}E^\beta{}_{\gamma}
\Phi^{\alpha(k-1)\gamma\dot\alpha(k-1)}\\
&&\qquad\qquad-
(k-1)\Phi_{\alpha(k)\dot\alpha(k-2)\dot\beta}E^{\dot\beta}{}_{\dot\gamma}
\Phi^{\alpha(k)\dot\alpha(k-2)\dot\gamma}+h.c.].\nonumber
\end{eqnarray}
The Lagrangian is invariant under gauge transformations
\begin{eqnarray*}
\delta\Phi^{\alpha(k)\dot\alpha(k-1)}&=&D\xi^{\alpha(k)\dot\alpha(k-1)}
+e_\beta{}^{\dot\alpha}\eta^{\alpha(k)\beta\dot\alpha(k-2)}
+\epsilon_{k+\frac12}{\lambda}
e^\alpha{}_{\dot\beta}\xi^{\alpha(k-1)\dot\alpha(k-1)\dot\beta}
\\
\delta\Phi^{\alpha(k-1)\dot\alpha(k)}&=&D\xi^{\alpha(k-1)\dot\alpha(k)}
+e^\alpha{}_{\dot\beta}\eta^{\alpha(k-2)\dot\alpha(k)\dot\beta}
+\epsilon_{k+\frac12}{\lambda}e_\beta{}^{\dot\alpha}\xi^{\alpha(k-1)\beta\dot\alpha(k-1)},
\end{eqnarray*}
where $\epsilon_{k+\frac12}=\pm1$. Note that the above consideration
does not fix a sign of $\epsilon_{k+\frac12}.$ As in integer spin
case we can construct the gauge invariant curvatures
\begin{eqnarray*}
{\cal
F}^{\alpha(k)\dot\alpha(k-1)}&=&D\Phi^{\alpha(k)\dot\alpha(k-1)}
+\epsilon_{k+\frac12}{\lambda}
e^\alpha{}_{\dot\beta}\Phi^{\alpha(k-1)\dot\alpha(k-1)\dot\beta}
\\
{\cal
F}^{\alpha(k-1)\dot\alpha(k)}&=&D\Phi^{\alpha(k-1)\dot\alpha(k)}
+\epsilon_{k+\frac12}{\lambda}e_\beta{}^{\dot\alpha}\Phi^{\alpha(k-1)\beta\dot\alpha(k-1)}
\end{eqnarray*}
Then Lagrangian variation can be compactly written as follows
\begin{eqnarray*}
(-1)^k\delta{\cal L}_{k+\frac12}&=&-{\cal
F}_{\alpha(k-1)\beta\dot\alpha(k-1)}e^\beta{}_{\dot\beta}
\delta\Phi^{\alpha(k-1)\dot\alpha(k-1)\dot\beta}+h.c. \, .
\end{eqnarray*}

Both in bosonic and in fermionic cases the variation of the higher
spin Lagrangians is completely expressed in geometric terms.

\section{Minimal $N=1$ supermultipets}\label{Section2}
In this section, we present the minimal massless $N=1$
supermultiplets in $4D$ AdS.  In the next sections, they will play
the role of building blocks for construct the extended
supermultiplets.

\subsection{Higher superspins}
{\bf Supermultiplet $(k+1/2,k)$} contains two massless fields with
spin $k$ and spin $k+1/2$. They are described by the fields
$$
f^{\alpha(k-1)\dot\alpha(k-1)},\quad
\Omega^{\alpha(k)\dot\alpha(k-2)},\quad
\Omega^{\alpha(k-2)\dot\alpha(k)}
$$
and
$$
\Phi^{\alpha(k)\dot\alpha(k-1)},\quad
\Phi^{\alpha(k-1)\dot\alpha(k)}
$$
respectively. The corresponding supertransformations are written in
the form
\begin{eqnarray*}
\delta
f^{\alpha(k-1)\dot\alpha(k-1)}&=&\alpha\Phi^{\alpha(k-1)\beta\dot\alpha(k-1)}\zeta_\beta
-\bar\alpha\Phi^{\alpha(k-1)\dot\alpha(k-1)\dot\beta}\zeta_{\dot\beta}
\\
\delta\Phi^{\alpha(k)\dot\alpha(k-1)}&=&\beta\Omega^{\alpha(k)\dot\alpha(k-2)}\zeta^{\dot\alpha}
+\gamma f^{\alpha(k-1)\dot\alpha(k-1)}\zeta^{\alpha}
\\
\delta\Phi^{\alpha(k-1)\dot\alpha(k)}&=&\bar\beta\Omega^{\alpha(k-2)\dot\alpha(k)}\zeta^{\alpha}
+\bar\gamma f^{\alpha(k-1)\dot\alpha(k-1)}\zeta^{\dot\alpha},
\end{eqnarray*}
were $\alpha,\beta,\gamma$ are the complex parameters. Parameters of
the $N=1$ supertransformations $\zeta^\alpha,\zeta^{\dot\alpha}$
satisfy the relation
\begin{equation}\label{AdScondition}
D\zeta^\alpha=-\lambda
e^\alpha{}_{\dot\beta}\zeta^{\dot\beta},\qquad
D\zeta^{\dot\alpha}=-\lambda e_\beta{}^{\dot\alpha}\zeta^{\beta}.
\end{equation}
Note that here and further we do not introduce any
supertransformation for auxiliary field $\Omega$ since the
calculations are done up to equations of motion for it
(\ref{TorsHS}). Invariance of the Lagrangian, $\delta({\cal
L}_k+{\cal L}_{k+\frac12})=0$ requires the restrictions on the
coefficients
$$
\alpha=i\frac{(k-1)}{4}\bar\beta,\quad\gamma=\lambda\beta,\quad
\beta=\epsilon_{k+\frac12}\bar\beta,\quad \epsilon_{k+\frac12}=\pm1.
$$
The free complex parameter $\beta$ can be taken pure real or pure
imaginary. In AdS space it relates the sign of mass-like term for
fermionic field and parity of bosonic field. Two cases
$\epsilon_{k+\frac12}=+1/-1$ correspond to different $N=1$ massless
supermultiplets with parity-even/odd boson. To fix the parameter
$\beta$ one calculates the commutator of two supertransformations on
bosonic field
\begin{eqnarray}\label{CommHS}
\frac{1}{\rho}{[\delta_1,\delta_2]}f^{\alpha(k-1)\dot\alpha(k-1)}&=&
\Omega^{\alpha(k-1)\beta\dot\alpha(k-2)}\xi_\beta{}^{\dot\alpha}+
\Omega^{\alpha(k-2)\dot\alpha(k-1)\dot\beta}\xi^\alpha{}_{\dot\beta}\nonumber\\
&& +\lambda(
f^{\alpha(k-2)\beta\dot\alpha(k-1)}\eta^{\alpha}{}_\beta+
f^{\alpha(k-1)\dot\alpha(k-2)\dot\beta}\eta^{\dot\alpha}{}_{\dot\beta}),
\end{eqnarray}
where
\begin{equation}\label{Parameters}
\xi_\beta{}^{\dot\alpha}=i(\zeta_1^{\dot\alpha}\zeta_2{}_\beta-\zeta_2^{\dot\alpha}\zeta_1{}_\beta),\quad
\eta^{\alpha}{}_\beta=i(\zeta_1^{\alpha}\zeta_2{}_\beta-\zeta_2^{\alpha}\zeta_1{}_\beta)
\end{equation}
and
$$
\rho=\frac{(k-1)}{4}\bar\beta\beta.
$$
We see that the commutator of these supertransformations is
combination of translation with parameter $\xi^{\alpha\dot\alpha}$
and Lorentz rotation with parameter $\eta^{\alpha\beta}$,
$\eta^{\dot\alpha\dot\beta}$ . It means that two corresponding
supercharges $Q_\alpha$, $Q_{\dot\alpha}$ satisfy the commutation
relations of $N=1$, AdS superalgebra
\begin{eqnarray*}
\{Q_\alpha, Q_{\dot\beta}\}&\sim&P_{\alpha\dot\beta},
\\
\{Q_\alpha, Q_{\beta}\}&\sim&\lambda M_{\alpha\beta},
\\
\{Q_{\dot\alpha}, Q_{\dot\beta}\}&\sim&\lambda
M_{\dot\alpha\dot\beta},
\end{eqnarray*}
where $P_{\alpha\dot\alpha},M_{\alpha(2)},M_{\dot\alpha(2)}$ are the
AdS generators.
\\
\\
{\bf Supermultiplet $(k,k-1/2)$} contains massless integer spin $k$
and half-integer spin $k-1/2$. Corresponding fields are
$$
f^{\alpha(k-1)\dot\alpha(k-1)},\quad
\Omega^{\alpha(k)\dot\alpha(k-2)},\quad
\Omega^{\alpha(k-2)\dot\alpha(k)}
$$
and
$$
\Phi^{\alpha(k-1)\dot\alpha(k-2)},\quad
\Phi^{\alpha(k-2)\dot\alpha(k-1)}.
$$
Supertransformations under the equations of motion for auxiliary
field $\Omega$ (\ref{TorsHS}) can be written
\begin{eqnarray*}
\delta
f^{\alpha(k-1)\dot\alpha(k-1)}&=&\alpha'\Phi^{\alpha(k-1)\dot\alpha(k-2)}\zeta^{\dot\alpha}
-\bar\alpha'\Phi^{\alpha(k-2)\dot\alpha(k-1)}\zeta^{\alpha},
\\
\delta\Psi^{\alpha(k-1)\dot\alpha(k-2)}&=&\beta'\Omega^{\alpha(k-1)\beta\dot\alpha(k-2)}\zeta_\beta
+\gamma'f^{\alpha(k-1)\dot\alpha(k-2)\dot\beta}\zeta_{\dot\beta},
\\
\delta\Psi^{\alpha(k-2)\dot\alpha(k-1)}&=&\bar\beta'\Omega^{\alpha(k-2)\dot\alpha(k-1)\dot\beta}
\zeta_{\dot\beta}
+\bar\gamma'f^{\alpha(k-2)\beta\dot\alpha(k-1)}\zeta_\beta.
\end{eqnarray*}
Lagrangian invariance $\delta({\cal L}_k+{\cal L}_{k-\frac12})=0$
gives
$$
\alpha'=\frac{i}{4(k-1)}\bar\beta',\quad \gamma'=\lambda\beta',\quad
\beta'=\epsilon_{k-\frac12}\bar\beta',\quad
\epsilon_{k-\frac12}=\pm1.
$$
Again free parameter $\beta'$ can be pure real/imaginary. It
corresponds to two different $N=1$ massless supermultiplets with
parity even/odd boson. Calculating commutator of two
supertransformations which is equal (\ref{CommHS}) we fix $\beta'$
$$
\rho=\frac{1}{4(k-1)}\bar\beta'\beta'.
$$

\subsection{Low superspins}
{\bf Supermultiplet $(1,3/2)$} contains the massless field with spin
3/2 which is described by 1-forms $\Phi^\alpha$, $\Phi^{\dot\alpha}$
with the Lagrangian (\ref{FerLag}) at $k=1$
\begin{eqnarray*}
{\cal L}_{\frac32}&=&-\Psi_{\beta}e^\beta{}_{\dot\beta}
D\Psi^{\dot\beta}-\epsilon_{\frac32}\lambda[
\Psi_{\beta}E^\beta{}_{\gamma} \Psi^{\gamma}+h.c.].
\end{eqnarray*}
Massless spin 1 is described by dynamical 1-form $f$ and auxiliary
0-forms $W^{\alpha(2)},W^{\dot\alpha(2)}$. The corresponding
Lagrangian looks like
\begin{eqnarray*}
\frac{1}{i}{\cal L}_1 &=&2E W_{\alpha(2)} W^{\alpha(2)} +
E_{\alpha(2)}W^{\alpha(2)} Df + h.c.
\end{eqnarray*}
It is evident that the Lagrangian is invariant under the gauge
transformations
\begin{eqnarray*}
\delta f&=&D\xi,\qquad \delta W^{\alpha(2)}=0.
\end{eqnarray*}
We do not introduce any supertransformation for auxiliary field
$W^{\alpha(2)}$ since the calculations are done up to equations of
motion which are equivalent to condition
\begin{eqnarray}\label{Tors1}
{\cal T} &=& Df + 2(E_{\alpha(2)} W^{\alpha(2)} +
E_{\dot\alpha(2)}W^{\dot\alpha(2)})\approx 0.
\end{eqnarray}
As an consequence of the above condition we have the relation
$$
E_{\alpha(2)}D{ W}^{\alpha(2)}+ E_{\dot\alpha(2)}D {
W}^{\dot\alpha(2)} \approx 0.
$$
Then the supertransformations can be rewritten in the form
\begin{eqnarray*}
\delta f &=& \alpha \Phi^\alpha\zeta_{\alpha} - \bar\alpha
\Phi^{\dot\alpha} \zeta_{\dot\alpha},
\\
\delta \Phi^\alpha &=& \beta e_{\beta\dot\beta} W^{\alpha\beta}
\zeta^{\dot\beta}+\gamma f\zeta^\alpha,
\\
\delta \Phi^{\dot\alpha} &=& \bar\beta e_{\beta\dot\beta}
W^{\dot\alpha\dot\beta} \zeta^{\beta}+\bar\gamma
f\zeta^{\dot\alpha}.
\end{eqnarray*}
Condition of the Lagrangian invariance $\delta({\cal L}_1+{\cal
L}_{\frac32})=0$ under the equation ${\cal T}\approx0$ yields
$$
\alpha=-i\frac{\bar\beta}{2},\quad
\gamma=-\frac{\lambda}{2}\beta,\quad
\beta=\epsilon_{\frac32}\bar\beta,\quad \epsilon_{\frac32}=\pm1.
$$
Commutator of two supertransformations has the form
\begin{eqnarray}\label{Comm1}
\frac{1}{\rho}{[\delta_1,\delta_2]}f&=& -2e_{\beta\dot\beta}
(W^{\alpha\beta}\xi_\alpha{}^{\dot\beta} +
W^{\dot\alpha\dot\beta}\xi^\beta{}_{\dot\alpha}),
\end{eqnarray}
where $\xi^{\alpha\dot\alpha}$ is the same as in (\ref{Parameters})
and $\rho=\frac{\bar\beta\beta}{4}$.
\\
\\
{\bf Supermultiplet $(1,1/2)$} contains the massless spin 1
described in the same way as in the previous case and the massless
spin 1/2 described by 0-forms $Y^\alpha$, $Y^{\dot\alpha}$. The
Lagrangian for spin 1/2 field has the form
\begin{eqnarray*}
{\cal L} &=& - Y_\alpha E^\alpha{}_{\dot\alpha} DY^{\dot\alpha}.
\end{eqnarray*}
Note that unlike the higher spin fermionic fields, there is no
mass-like term in the above Lagrangian. Supertransformations, up to
the equations of motion for auxiliary field $W^{\alpha\beta}$, are
written as follows
\begin{eqnarray*}
\delta f &=& \alpha' e_{\alpha\dot\alpha}Y^\alpha\zeta^{\dot\alpha}
- \bar\alpha' e_{\alpha\dot\alpha}Y^{\dot\alpha} \zeta^{\alpha},
\\
\delta Y^\alpha &=& \beta' W^{\alpha\beta} \zeta_\beta,
\\
\delta Y^{\dot\alpha} &=& \bar\beta' W^{\dot\alpha\dot\beta}
\zeta_{\dot\beta}.
\end{eqnarray*}
Lagrangian invariance $\delta({\cal L}_1+{\cal L}_{\frac12})=0$
under the equations ${\cal T}\approx0$ (\ref{Tors1}) yields
$$
\alpha'=-\frac{i}{4}\bar\beta'.
$$
Calculating the commutator of the superetransformations leads to
relations (\ref{Comm1}) and allows to fix the parameter
$\rho=\frac{\bar\beta'\beta'}{8}$.
\\
\\
{\bf Supermultiplet $(0,1/2)$} contains the massless spin 1/2 and
one massless spin 0 which is described by the dynamical 0-form $W$
and the auxiliary 0-form $W^{\alpha\dot\alpha}$. The Lagrangian for
spin 0 has the form
\begin{eqnarray*}
\frac{1}{i}{\cal L} &=& - \frac12E W_{\alpha\dot\alpha}
W^{\alpha\dot\alpha} - E_{\alpha\dot\alpha} W^{\alpha\dot\alpha}
DW+2\lambda^2 E W^2.
\end{eqnarray*}
Using the equation of motion for auxiliary field
$W^{\alpha\dot\alpha}$
\begin{eqnarray*}
{\cal W} &=& DW + e_{\alpha\dot\alpha} W^{\alpha\dot\alpha}\approx 0
\end{eqnarray*}
one gets the  relation
\begin{eqnarray*}
e_{\alpha\dot\alpha} D{ W}^{\alpha\dot\alpha} \approx 0
\quad\Rightarrow\quad E^{\alpha}{}_{\dot\gamma} D{
W}^{\beta\dot\gamma} \approx \frac12 \varepsilon^{\alpha\beta}
E_{\gamma}{}_{\dot\gamma} D{ W}^{\gamma\dot\gamma}.
\end{eqnarray*}
We use the following anzac for supertransformations
\begin{eqnarray*}
\delta W &=& \alpha_0 Y^\alpha \zeta_{\alpha} - \bar{\alpha}_0
Y^{\dot\alpha} \zeta_{\dot\alpha},
\\
\delta Y^\alpha &=& \beta_0 W^{\alpha\dot\alpha} \zeta_{\dot\alpha}
+ \gamma_0 W \zeta^\alpha,
\\
\delta Y^{\dot\alpha} &=& \bar\beta_0 W^{\alpha\dot\alpha}
\zeta_{\alpha}  + \bar\gamma_0 W \zeta^{\dot\alpha}
\end{eqnarray*}
with a set of arbitrary complex parameters $\alpha_0, \beta_0,
\gamma_0$. Invariance of the Lagrangian $\delta({\cal L}_0+{\cal
L}_{\frac12})=0$ under the equation ${\cal W}\approx0$ gives
restrictions on the parameters
$$
\alpha_0=\frac{i}{2}\bar\beta_0,\quad \gamma_0=\lambda\beta_0.
$$
Commutator of two supertransformations for spin 0 field has the form
\begin{eqnarray*}
\frac{1}{\rho}{[\delta_1,\delta_2]}W&=&W^{\alpha\dot\alpha}\xi_{\alpha\dot\alpha},
\end{eqnarray*}
where $\xi_{\alpha\dot\alpha}$ is the parameter of translation
(\ref{Parameters}) and $\rho=\bar\beta_0\beta_0$. Parameter $\beta$
can be taken purely real/imaginary depending on supermultiplet with
parity even/odd spin 0 field.
\\
\\
{\bf Chiral supermultiplet $(0_+,0_-,1/2)$} contains one massless
spin 1/2 and two massless parity even/odd spins $0_+/0_-$. In this
case the spin 0 is described by the complex scalar field $W$. The
corresponding supertransformations have form
\begin{align}
& \delta W = 2\alpha_0 Y^\alpha \zeta_{\alpha} && \delta Y^\alpha =
\beta_0 W^{\alpha\dot\alpha} \zeta_{\dot\alpha} + \gamma_0W
\zeta^\alpha
\\
& \delta \bar W = -2\bar\alpha_0 Y^{\dot\alpha} \zeta_{\dot\alpha}
&& \delta Y^{\dot\alpha} = \bar\beta_0 \bar{W}^{\alpha\dot\alpha}
\zeta_{\alpha}  + \bar\gamma_0 \bar{W} \zeta^{\dot\alpha},
\end{align}
where
$$
\alpha_0=\frac{i}{2}\bar\beta_0,\quad \gamma_0=\lambda\beta_0.
$$
The commutators of the supertransformations are written as follows
\begin{eqnarray*}
\frac{1}{\rho}{[\delta_1,\delta_2]}W&=&
W^{\alpha\dot\alpha}\xi_{\alpha\dot\alpha} ,\qquad
\frac{1}{\rho}{[\delta_1,\delta_2]}\bar{W}=
\bar{W}^{\alpha\dot\alpha}\xi_{\alpha\dot\alpha},
\end{eqnarray*}
where $\rho=\bar\beta_0\beta_0$. If to redenote the complex field in
the form
$$
W=W_++iW_-\qquad
W^{\alpha\dot\alpha}=W_+^{\alpha\dot\alpha}+iW_-^{\alpha\dot\alpha}
$$
$$
\bar{W}=W_+-iW_-\qquad
\bar{W}^{\alpha\dot\alpha}=W_+^{\alpha\dot\alpha}-iW_-^{\alpha\dot\alpha},
$$
then at real $\beta_0$ the field $W_+$ is parity even spin $0$ field
while $W_-$ is parity odd one. At imaginary $\beta_0$ the $W_+$ is
parity odd field and $W_-$ is parity even one.

\section{$N$-extended supermultiplets}\label{Section3}
In this section, we consider the massless $N$-extended higher spin
supermultiplets in $4D$ AdS. As we pointed out in the Introduction,
for given maximal integer or half-integer spin $k$ in the
supermultiplt, the parameter $N$ satisfies the relation $N\leq4k$.
For each spin $k$ we describe the field contents and the
corresponding field variables. Then, we derive the
supertransformations and show that the specially defined free
Lagrangians are invariant under these transformations. Finally, we
prove that the constructed supertransformations form the on-shell
closed $N$- extended $4D$ AdS superalgebras.

\subsection{$N\leq2k-3$}
In this case, the massless supermultiplets contain the massless
fields with spins
$$
k,k-\frac12,k-1,...,k-\frac{N-1}{2},k-\frac{N}{2},
$$
where $k$ is an arbitrary integer or half-integer. We will write it
compactly as
$$
k-\frac{m}{2},\quad m=0,1,...,N.
$$
The number of massless fields with the given spin $k-\frac{m}{2}$ is
equal to $\frac{N!}{m!(N-m)!}$. One can see that the minimal spin
equals to $\frac32$ in the boundary case $N=2k-3$. So all massless
fields entering extended supermultiplets are uniformly described in
the Section 1.

Let us introduce the bosonic field variables
$$
f_{k-\frac{m}{2},i[m]}{}^{\alpha(k-\frac{m+2}{2})\dot\alpha(k-\frac{m+2}{2})},\quad
\Omega_{k-\frac{m}{2},i[m]}{}^{\alpha(k-\frac{m}{2})\dot\alpha(k-\frac{m+4}{2})}
$$
and fermionic ones
$$
\Phi_{k-\frac{m}{2},i[m]}{}^{\alpha(k-\frac{m+1}{2})\dot\alpha(k-\frac{m+3}{2})}.
$$
Here the first lower index denotes the spin of the field, the
compact index $i[m]=[i_1i_2...i_m]$ denotes the antisymmetric
combination of indices $i=1,2,...N$ and corresponds to antisymmetric
representation of the internal symmetry group $SO(N)$. If the
maximal spin $k$ is integer then $m$ takes even values
$0,2,...,2[\frac{N}{2}]$ for bosonic fields and odd values
$1,3,...,2[\frac{N-1}{2}]+1$ for fermionic ones. In the case maximal
half-integer spin $k$ the parameter m takes the even values for
fermions and odd ones for bosons. .

The generic anzac for the linear supertransformations is chosen in
the following form with a set of arbitrary complex coefficients
$\alpha_m, \alpha'_m, \beta_m, \beta'_m, \gamma_m, \gamma'_m$
\begin{eqnarray}
\delta
f_{k-\frac{m}{2},i[m]}{}^{\alpha(k-\frac{m+2}{2})\dot\alpha(k-\frac{m+2}{2})}&=&
\alpha'_{m}\Phi_{k-\frac{m+1}{2},i[m]j}{}^{\alpha(k-\frac{m+2}{2})\dot\alpha(k-\frac{m+4}{2})}\zeta^{j}{}^{\dot\alpha}\nonumber\\
&&-\bar\alpha'_{m}\Phi_{k-\frac{m+1}{2},i[m]j}{}^{\alpha(k-\frac{m+4}{2})\dot\alpha(k-\frac{m+2}{2})}\zeta^{j}{}^{\alpha}\nonumber\\
&&+{\alpha_{m}}\Phi_{k-\frac{m-1}{2},i[m-1]}{}^{\alpha(k-\frac{m+2}{2})\beta\dot\alpha(k-\frac{m+2}{2})}\zeta_{i}{}_{\beta}\nonumber\\
&&-{\bar\alpha_{m}}\Phi_{k-\frac{m-1}{2},i[m-1]}{}^{\alpha(k-\frac{m+2}{2})\dot\alpha(k-\frac{m+2}{2})\dot\beta}\zeta_{i}{}_{\dot\beta}
\label{STB}\\
\delta\Phi_{k-\frac{m}{2},i[m]}{}^{\alpha(k-\frac{m+1}{2})\dot\alpha(k-\frac{m+3}{2})}&=&
\beta_{m}\Omega_{k-\frac{m+1}{2},i[m]j}{}^{\alpha(k-\frac{m+1}{2})\dot\alpha(k-\frac{m+5}{2})}\zeta^{j,\dot\alpha}\nonumber\\
&&
+{\beta'_{m}}\Omega_{k-\frac{m-1}{2},i[m-1]}{}^{\alpha(k-\frac{m+1}{2})\beta\dot\alpha(k-\frac{m+3}{2})}\zeta_{i,\beta}\nonumber\\
&&+\gamma_{m}f_{k-\frac{m+1}{2},i[m]j}{}^{\alpha(k-\frac{m+3}{2})\dot\alpha(k-\frac{m+3}{2})}\zeta^{j,\alpha}\nonumber\\
&&
+{\gamma'_{m}}f_{k-\frac{m-1}{2},i[m-1]}{}^{\alpha(k-\frac{m+1}{2})\dot\alpha(k-\frac{m+3}{2})
\dot\beta}\zeta_{i,\dot\beta}.\label{STF}
\end{eqnarray}
Here $\zeta_i^{\alpha},\zeta_i^{\dot\alpha}$ are parameters of
extended supertransformations satisfying the  conditions
(\ref{AdScondition}). Lagrangian is defined as ${\cal L}=\sum_m{\cal
L}_{k-\frac{m}{2}},$ where ${\cal L}_{k-\frac{m}{2}}$ is the
Lagrangian for free field with spin $k-\frac{m}{2}$. Invariance of
the Lagrangian under these supertransformations leads to
restrictions on the arbitrary parameters
$$
\alpha_m=\frac{i(k-\frac{m+2}{2})}{4}\bar\beta_{m-1}\quad
\gamma_m=\lambda\beta_m,\quad
\beta_m=\epsilon_{k-\frac{m}{2}}\bar\beta_m
$$
$$
\alpha'_m=\frac{i}{4(k-\frac{m+2}{2})}\bar\beta'_{m+1}\quad
\gamma'_m=\lambda\beta'_m,\quad
\beta'_m=\epsilon_{k-\frac{m}{2}}\bar\beta'_m.
$$
In these relations $\epsilon_{k-\frac{m}{2}}=+1$ or
$\epsilon_{k-\frac{m}{2}}=-1$ for any $m$ depending on parity of the
corresponding field. It means that there are two families of
parameters $\beta_{m}$ and $\beta'_{m}$. In order to relate them
with each other we require the closer of the supertransformations
algebra (\ref{STB}), (\ref{STF}). It yields the condition
\begin{equation}\label{ClosedHS}
{\alpha_{m}}\beta_{m-1}-\alpha'_{m}{\beta'_{m+1}}=0.
\end{equation}

Calculation of the commutator for the two supertransformations
(\ref{STB}) allows to obtain the result
\begin{eqnarray}\label{CommHSN}
\frac{1}{\rho}{[\delta_1,\delta_2]}
f_{k-\frac{m}{2},i[m]}{}^{\alpha(k-\frac{m+2}{2})\dot\alpha(k-\frac{m+2}{2})}&=&
\Omega_{k-\frac{m}{2},i[m]}{}^{\alpha(k-\frac{m+2}{2})\beta\dot\alpha(k-\frac{m+4}{2})}\xi_\beta{}^{\dot\alpha}\nonumber\\
&&
+\Omega_{k-\frac{m}{2},i[m]}{}^{\alpha(k-\frac{m+4}{2})\dot\alpha(k-\frac{m+2}{2})\dot\beta}\xi^\alpha{}_{\dot\beta}\nonumber\\
&& +\lambda(
f_{k-\frac{m}{2},i[m]}{}^{\alpha(k-\frac{m+2}{2})\dot\alpha(k-\frac{m+4}{2})\dot\beta}
\eta_{\dot\beta}{}^{\dot\alpha}\\
&&+f_{k-\frac{m}{2},i[m]}{}^{\alpha(k-\frac{m+4}{2})\beta\dot\alpha(k-\frac{m+2}{2})}
\eta_{\beta}{}^{\alpha})\nonumber
\\
&& +\lambda
f_{k-\frac{m}{2},i[m-1]j}{}^{\alpha(k-\frac{m+2}{2})\dot\alpha(k-\frac{m+2}{2})}z^j{}_i,
\nonumber
\end{eqnarray}
where
\begin{eqnarray}
&&
\xi_\beta{}^{\dot\alpha}=i(\zeta_1{}_{j,\beta}\zeta_2^{j}{}^{\dot\alpha}-\zeta_2{}_{j,\beta}\zeta_1^{j}{}^{\dot\alpha}),\quad
\eta_{\dot\beta}{}^{\dot\alpha}=i(\zeta_1{}_{j,\dot\beta}\zeta_2^{j}{}^{\dot\alpha}-\zeta_2{}_{j,\dot\beta}\zeta_1^{j}{}^{\dot\alpha})
\label{SpaceParam}
\\
&&
z^j{}_i=i(\zeta_1^{j,\beta}\zeta_2{}_{i}{}_{\beta}-\zeta_2^{j,\beta}\zeta_1{}_{i}{}_{\beta}
+\zeta_1^{j,\dot\beta}\zeta_2{}_{i}{}_{\dot\beta}-\zeta_2^{j,\dot\beta}\zeta_1{}_{i}{}_{\dot\beta}),\quad
z^{ij}=-z^{ji}.\label{InternParam}
\end{eqnarray}
One can see that commutator (\ref{CommHSN}) is equal to combinations
of translations, Lorentz rotations and internal $SO(N)$
transformations with parameters $\xi^{\alpha\dot\alpha}$,
$\eta^{\alpha\beta}$ and $z^{ij}$ respectively\footnote{We have
calculated such a commutator only for bosonic fields. For fermionic
fields the result will  be the same, however the computations became
to be more complicated and tedious}. From (\ref{ClosedHS}),
(\ref{CommHSN}) we have restriction for the parameters
$$
\bar\beta_{m-1}\beta_{m-1}=\frac{4\rho}{(k-\frac{m+2}{2})},\quad
\bar\beta'_{m+1}\beta'_{m+1}=4\rho(k-\frac{m+2}{2}).
$$
The form of the above commutator shows to prove that the
supercharges $Q_\alpha{}^i$, $Q_{\dot\alpha}{}^i$ corresponding to
the supertransformations (\ref{STB}), (\ref{STF}) satisfy the
commutation relations of the extended AdS superalgebra
\begin{eqnarray}
\label{Q} \{Q^i_\alpha,
Q^j_{\dot\beta}\}&\sim&\delta^{ij}P_{\alpha\dot\beta},\nonumber
\\
\{Q^i_\alpha,
Q^j_{\beta}\}&\sim&\lambda(\delta^{ij}M_{\alpha\beta}+\frac12\varepsilon_{\alpha\beta}T^{ij}),
\\
\{Q^i_{\dot\alpha},
Q^j_{\dot\beta}\}&\sim&\lambda(\delta^{ij}M_{\dot\alpha\dot\beta}+\frac12\varepsilon_
{\dot\alpha\dot\beta}T^{ij}) \nonumber,
\end{eqnarray}
where $P_{\alpha\dot\alpha},M_{\alpha(2)},M_{\dot\alpha(2)}$ are the
AdS generators and $T^{ij}=-T^{ji}$ are the generators of internal
$SO(N)$ group symmetry.

\subsection{$2k-3<N\leq2k$}
In order to go beyond $N>2k-3$ we should include the massless fields
with lower spins to the supermultiplets. In the case $N=2k-2$ it is
sufficient to add the massless spin 1 and the corresponding fields
$$
f_{1,i[2k-2]},\quad W_{i[2k-2]}{}^{\alpha(2)}.
$$
Analogically, in the cases $N=2k-1$ and $N=2k$ we also should add
the massless fields with spins $\frac12$
$$
Y_{i[2k-1]}{}^\alpha,\quad Y_{i[2k-1]}{}^{\dot\alpha}
$$
and set of complex fields for spins 0
$$
W_{i[2k]},\quad W_{i[2k]}{}^{\alpha\dot\alpha}\qquad
\bar{W}_{i[2k]},\quad \bar{W}_{i[2k]}{}^{\alpha\dot\alpha}.
$$
I this case the anzac for supertransformations with a set of
arbitrary parameters looks like
\begin{eqnarray}
\delta\Phi_{\frac{3}{2},i[2k-3]}{}^{\alpha}&=&
{\beta'_{2k-3}}\Omega_{2,i[2k-4]}{}^{\alpha\beta}\zeta_{i,\beta}
+{\gamma'_{2k-3}}f_{2,i[2k-4]}{}^{\alpha\dot\beta}\zeta_{i,\dot\beta}\nonumber\\
&&+ \beta_{2k-3}e_{\beta\dot\beta} W_{i[2k-3]j}{}^{\alpha\beta}
\zeta^j{}^{\dot\beta}+\gamma_{2k-3}f_{1,i[2k-3]j}\zeta^j{}^\alpha,\label{ST3/2}
\\
\delta f_{1,i[2k-2]} &=& {\alpha_{2k-2}}
\Phi_{\frac32,i[2k-3]}{}^\alpha\zeta_{i}{}_{\alpha} +\alpha'_{2k-2}
e_{\alpha\dot\alpha}Y_{i[2k-2]j}{}^\alpha\zeta^j{}^{\dot\alpha},\label{ST1}
\\
\delta Y_{i[2k-1]}{}^\alpha &=& {\beta'_{2k-1}}
W_{i[2k-2]}{}^{\alpha\beta} \zeta_i{}_\beta
\nonumber\\
&&+\beta_{2k-1} W_{i[2k-1]j}{}^{\alpha\dot\alpha}
\zeta^j{}_{\dot\alpha} + \gamma_{2k-1} W_{i[2k-1]j}
\zeta^j{}^\alpha,\label{ST1/2}
\\
\delta W_{i[2k]} &=& {2\alpha_{2k}} Y_{i[2k-1]}{}^\alpha
\zeta_i{}_{\alpha},\quad \delta \bar W_{i[2k]} = -{2\bar\alpha_{2k}}
Y_{i[2k-1]}{}^{\dot\alpha} \zeta_i{}_{\dot\alpha}.\label{ST0}
\end{eqnarray}
Lagrangian is defined as a sum of Lagrangians for all fields in the
supermultiplet. Invariance of the Lagrangian yields restrictions for
the arbitrary parameters
$$
\alpha_{2k-2}=-i\frac{\bar\beta_{2k-3}}{2},\quad
\gamma_{2k-3}=-\frac{\lambda}{2}\beta_{2k-3},\quad
\beta_{2k-3}=\epsilon_{\frac32}\bar\beta_{2k-3},\quad
\epsilon_{\frac32}=\pm1,
$$
$$
\alpha'_{2k-2}=-\frac{i}{4}\bar\beta'_{2k-1},\quad
\alpha_{2k}=\frac{i}{2}\bar\beta_{2k-1},\quad
\gamma_{2k-1}=\lambda\beta_{2k-1}.
$$
As a result, there are three arbitrary complex parameters
$\beta_{2k-3}$, $\beta'_{2k-1}$ and $\beta_{2k-1}$ which can be
either purely real or purely imaginary depending on even or odd
parity of bosonic fields entering supermultiplets. We fix them by
the requirement that commutators for spin 1 and spin 0 are closed.
It gives the condition
\begin{equation}\label{Closed1}
{\alpha_{2k-2}}\beta_{2k-3}-\alpha'_{2k-2}{\beta'_{2k-1}}=0.
\end{equation}
Then the commutators for the spin 1 field has the following form up
to the gauge transformations
\begin{eqnarray}
\frac{1}{\rho}{[\delta_1,\delta_2]}f_{i[2k-2]}&=&
-2e_{\alpha\dot\alpha}(W_{i[2k-2]}{}^{\alpha\beta}\xi_\beta{}^{\dot\alpha}+
W_{i[2k-2]}{}^{\dot\alpha\dot\beta}\xi^\alpha{}_{\dot\beta})
+\lambda f_{i[2k-3]j}z^j{}_i,\label{Comm1N}
\\
\frac{1}{\rho}{[\delta_1,\delta_2]}W_{i[2k]} &=&
 W_{i[2k]}{}^{\alpha\dot\alpha}\xi_{\alpha\dot\alpha},\label{Comm0N}
\end{eqnarray}
where $\xi^{\alpha\dot\alpha}$ and $z^{ij}$ are parameters of
translations and internal $SO(N)$ symmetry defined by
(\ref{SpaceParam}) and (\ref{InternParam}). The relations
(\ref{Closed1}), (\ref{Comm1N}), (\ref{Comm0N}) lead to the
conditions for the parameters
$$
\bar\beta_{2k-3}\beta_{2k-3}=4\rho,\quad
\bar\beta'_{2k-1}\beta'_{2k-1}=8\rho,\quad
\bar\beta_{2k-1}\beta_{2k-1}=\rho.
$$
Using the commutators (\ref{Comm1N}), (\ref{Comm0N}) we can show
that the corresponding supercharges satisfy the relations (\ref{Q}).

\subsection{$2k<N<4k$}\label{Sec4.3}
To extend supersymmetry further, i.e. to consider the case $N>2k$ we
should include into supermultiplets more massless fields with spins
$$
\frac12,1,\frac{m}{2}-k\quad m=2k+3,2k+4,...,N-1,N.
$$
We introduce the additional field variables for spin $\frac12$
$$
Y_{i[2k+1]}{}^\alpha,\quad Y_{i[2k+1]}{}^{\dot\alpha},
$$
for spin 1
$$
f_{1,i[2k+2]},\quad W_{i[2k+2]}{}^{\alpha(2)},
$$
and for higher spins
$$
f_{\frac{m}{2}-k,i[m]}{}^{\alpha(\frac{m-2}{2}-k)\dot\alpha(\frac{m-2}{2}-k)},\quad
\Omega_{\frac{m}{2}-k,i[m]}{}^{\alpha(\frac{m}{2}-k)\dot\alpha(\frac{m-4}{2}-k)},\quad
m=2k+4,...,2[\frac{N}{2}]
$$
$$
\Phi_{\frac{m}{2}-k,i[m]}{}^{\alpha(\frac{m-1}{2}-k)\dot\alpha(\frac{m-3}{2}-k)},\quad
m=2k+3,...,2[\frac{N-1}{2}]+1
$$
Additional anzac for the supertransformations is chosen in the
following form with a set of arbitrary parameters
\begin{eqnarray*}
\delta
f_{\frac{m}{2}-k,i[m]}{}^{\alpha(\frac{m-2}{2}-k)\dot\alpha(\frac{m-2}{2}-k)}&=&
\alpha'_{m}\Phi_{\frac{m+1}{2}-k,i[m]j}{}^{\alpha(\frac{m-2}{2}-k)\beta\dot\alpha(\frac{m-2}{2}-k)}\zeta^{j}{}_{\beta}\\
&&+{\alpha_{m}}\Phi_{\frac{m-1}{2}-k,i[m-1]}{}^{\alpha(\frac{m-2}{2}-k)\dot\alpha(\frac{m-4}{2}-k)}\zeta_{i}{}^{\dot\alpha}
+h.c.\quad m\geq2k+4,\\
\delta\Phi_{\frac{m}{2}-k,i[m]}{}^{\alpha(\frac{m-1}{2}-k)\dot\alpha(\frac{m-3}{2}-k)}&=&
\beta_{m}\Omega_{\frac{m+1}{2}-k,i[m]j}{}^{\alpha(\frac{m-1}{2}-k)\beta\dot\alpha(\frac{m-3}{2}-k)}
\zeta^{j}{}_\beta\\
&&
+{\beta'_{m}}\Omega_{\frac{m-1}{2}-k,i[m-1]}{}^{\alpha(\frac{m-1}{2}-k)\dot\alpha(\frac{m-5}{2}-k)}\zeta_{i}{}^{\dot\alpha}\\
&&+\gamma_{m}f_{\frac{m+1}{2}-k,i[m]j}{}^{\alpha(\frac{m-1}{2}-k)\dot\alpha(\frac{m-3}{2}-k)
\dot\beta}\zeta^{j}{}_{\dot\beta}\\
&&
+{\gamma'_{m}}f_{\frac{m-1}{2}-k,i[m]}{}^{\alpha(\frac{m-3}{2}-k)\dot\alpha(\frac{m-3}{2}-k)}
\zeta_{i}{}^{\alpha} \quad m\geq2k+5,
\\
\delta\Phi_{\frac{3}{2},i[2k+3]}{}^{\alpha}&=&
{\beta_{2k+3}}\Omega_{2,i[2k+3]j}{}^{\alpha\beta}\zeta^j{}_{\beta}
+{\gamma_{2k+3}}f_{2,i[2k+3]j}{}^{\alpha\dot\beta}\zeta^j{}_{\dot\beta}\\
&&+ \beta'_{2k+3}e_{\beta\dot\beta} W_{i[2k+2]}{}^{\alpha\beta}
\zeta_i{}^{\dot\beta}+\gamma'_{2k+3}f_{1,i[2k+2]}\zeta_i{}^\alpha,
\\
\delta f_{1,i[2k+2]} &=& {\alpha'_{2k+2}}
\Phi_{\frac32,i[2k+2]j}{}^\alpha\zeta^{j}{}_{\alpha} +\alpha_{2k+2}
e_{\alpha\dot\alpha}Y_{i[2k+1]}{}^\alpha\zeta_i{}^{\dot\alpha},
\\
\delta Y_{i[2k+1]}{}^\alpha &=& {\beta_{2k+1}}
W_{i[2k+1]j}{}^{\alpha\beta} \zeta^j{}_\beta
\\
&&+\beta'_{2k+1}\bar{ W}_{i[2k]}{}^{\alpha\dot\alpha}
\zeta_i{}_{\dot\alpha} + \gamma'_{2k+1} \bar{W}_{i[2k]}
\zeta_i{}^\alpha,
\\
\delta W_{i[2k]} &=& -{2\bar\alpha'_{2k}} Y_{i[2k]j}{}^{\dot\alpha}
\zeta^j{}_{\dot\alpha} \quad \delta \bar W_{i[2k]} = {2\alpha'_{2k}}
Y_{i[2k]j}{}^{\alpha} \zeta^j{}_{\alpha}.
\end{eqnarray*}
Lagrangian is defined as a sum of the Lagrangians for all fields in
the supermultiplet. Condition of invariance of the Lagrangian
invariance yields restrictions on the arbitrary parameters
$$
\alpha_m=\frac{i}{4(\frac{m-2}{2}-k)}\bar\beta_{m-1},\quad
\gamma_m=\lambda\beta_m,\quad
\beta_m=\epsilon_{\frac{m}{2}-k}\bar\beta_m,
$$
$$
\alpha'_m=\frac{i(\frac{m-2}{2}-k)}{4}\bar\beta'_{m+1},\quad
\gamma'_m=\lambda\beta'_m,\quad
\beta'_m=\epsilon_{\frac{m}{2}-k}\bar\beta'_m,
$$
$$
\alpha'_{2k+2}=-i\frac{\bar\beta'_{2k+3}}{2},\quad
\gamma'_{2k+3}=-\frac{\lambda}{2}\beta'_{2k+3},\quad
\beta'_{2k+3}=\epsilon_{\frac32}\bar\beta'_{2k+3},\quad
\epsilon_{\frac32}=\pm1,
$$
$$
\alpha_{2k+2}=-\frac{i}{4}\bar\beta_{2k+1},\quad
\alpha'_{2k}=\frac{i}{2}\bar\beta'_{2k+1},\quad
\gamma'_{2k+1}=\lambda\beta'_{2k+1}.
$$
To close algebra of the supertransformations we impose the
conditions
\begin{equation}\label{ClosedHS+}
{\alpha_{m}}\beta_{m-1}-\alpha'_{m}{\beta'_{m+1}}=0,\quad
\alpha_{2k+2}{\beta_{2k+1}} -{\alpha'_{2k+2}}\beta'_{2k+3}=0,\quad
{2\alpha_{2k}}\beta_{2k-1}+{2\bar\alpha'_{2k}}\bar\beta'_{2k+1}=0
\end{equation}
As a result, the commutators of supertransformations have the form
\begin{eqnarray}
\label{commutators}
 \frac{1}{\rho}{[\delta_1,\delta_2]}
f_{\frac{m}{2}-k,i[m]}{}^{\alpha(k-\frac{m+2}{2})\dot\alpha(k-\frac{m+2}{2})}&=&
\Omega_{\frac{m}{2}-k,i[m]}{}^{\alpha(k-\frac{m+2}{2})\beta\dot\alpha(k-\frac{m+4}{2})}
\xi_\beta{}^{\dot\alpha} \nonumber\\
&&
+\Omega_{\frac{m}{2}-k,i[m]}{}^{\alpha(k-\frac{m+4}{2})\dot\alpha(k-\frac{m+2}{2})\dot\beta}
\xi^\alpha{}_{\dot\beta} \nonumber\\
&& +\lambda(
f_{\frac{m}{2}-k,i[m]}{}^{\alpha(k-\frac{m+2}{2})\dot\alpha(k-\frac{m+4}{2})\dot\beta}
\eta_{\dot\beta}{}^{\dot\alpha} \\
&&+f_{\frac{m}{2}-k,i[m]}{}^{\alpha(k-\frac{m+4}{2})\beta\dot\alpha(k-\frac{m+2}{2})}
\eta_{\beta}{}^{\alpha}) \nonumber
\\
&& +\lambda
f_{\frac{m}{2}-k,i[m-1]j}{}^{\alpha(k-\frac{m+2}{2})\dot\alpha(k-\frac{m+2}{2})}z^j{}_i,
\nonumber
\\
\frac{1}{\rho}{[\delta_1,\delta_2]}f_{1,i[2k+2]}&=&
-2e_{\alpha\dot\alpha}(W_{i[2k+2]}{}^{\alpha\beta}\xi_\beta{}^{\dot\alpha}+
W_{i[2k+2]}{}^{\dot\alpha\dot\beta}\xi^\alpha{}_{\dot\beta})
+\lambda f_{i[2k-3]j}z^j{}_i, \nonumber
\\
\frac{1}{\rho}{[\delta_1,\delta_2]}W_{i[2k]} &=&
 W_{i[2k]}{}^{\alpha\dot\alpha}\xi_{\alpha\dot\alpha}
 +\lambda{W}_{i[2k-1]j}z^j{}_i, \nonumber
\end{eqnarray}
where $\xi^{\alpha\dot\alpha}$, $\eta^{\alpha\beta}$ and $z^{ij}$
are defined by (\ref{SpaceParam}) and (\ref{InternParam}). Form of
the above commutators and relation (\ref{ClosedHS+}) allows to find
the additional restrictions for the arbitrary parameters
$$
\bar\beta_{m-1}\beta_{m-1}=4\rho(\frac{m-2}{2}-k),\quad
\bar\beta'_{m+1}\beta'_{m+1}=\frac{4\rho}{(\frac{m-2}{2}-k)},
$$
$$
\bar\beta'_{2k+3}\beta'_{2k+3}=4\rho,\quad
\bar\beta_{2k+1}\beta_{2k+1}=8\rho,\quad
\bar\beta'_{2k+1}\beta'_{2k+1}=\rho.
$$
The commutators (\ref{commutators}) allows to calculate the algebra
of supercharges which will have the form (\ref{Q}). As a result we
obtain the on-shell $N$-extended component free Lagrangian
formulations for the supermultiplets with $2k<N<4k$.

\subsection{$N=4k$}
Now we consider a special case of maximal $N$-extended supersymmetry
with the highest spin $k$ in the supermultiplet. Such a
supermultiplet contains the massless fields with all spins from $k$
to $0$. Field variables are the same as in the $N=2k$ case but now
$i=1,2...,4k$
$$
f_{k-\frac{m}{2},i[m]}{}^{\alpha(k-\frac{m+2}{2})\dot\alpha(k-\frac{m+2}{2})},\quad
\Omega_{k-\frac{m}{2},i[m]}{}^{\alpha(k-\frac{m}{2})\dot\alpha(k-\frac{m+4}{2})}
$$
for bosonic higher spin fields and
$$
\Phi_{k-\frac{m}{2},i[m]}{}^{\alpha(k-\frac{m+1}{2})\dot\alpha(k-\frac{m+3}{2})}
$$
for fermionic higher spin fields. For spin 1 we introduce the fields
$$
f_{1,i[2k-2]},\quad W_{i[2k-2]}{}^{\alpha(2)},
$$
for spin 1/2 we introduce the fields
$$
Y_{i[2k-1]}{}^\alpha,\quad Y_{i[2k-1]}{}^{\dot\alpha}.
$$
Besides, we introduce a set of complex fields for spin 0
$$
W_{i[2k]},\quad W_{i[2k]}{}^{\alpha\dot\alpha}\qquad
\bar{W}_{i[2k]},\quad \bar{W}_{i[2k]}{}^{\alpha\dot\alpha}
$$
which are subject to the condition
\begin{equation}\label{ScalCond}
 W_{i[2k]}=\frac{1}{(2k)!}{\cal
E}_{i[2k]}{}^{j[2k]}\bar W_{j[2k]},
\end{equation}
where $k$ is an arbitrary integer. As before all field variables are
totally antisymmetric over the indices $i=1,2,...N$. Totally
antisymmetric invariant tensors ${\cal E}_{i[4k]}={\cal
E}_{[i_1i_2...i_{4k}]}$ are normalized as follows ${\cal
E}_{12...4k}=1$.

The supertransformations in the case under consideration are the
same as in the case $N=2k$ (\ref{STB}), (\ref{STF}), (\ref{STB}),
(\ref{ST3/2}), (\ref{ST1}), (\ref{ST1/2}). While for the spin 0
components, the  supertransformations compatible with
(\ref{ScalCond}) look like
\begin{eqnarray}\label{ST04k}
\delta W_{i[2k]} &=& {2\alpha_{2k}} Y_{i[2k-1]}{}^\alpha
\zeta_i{}_{\alpha} -\frac{2\bar\alpha_{2k}}{(2k-1)!}{\cal
E}_{i[2k]}{}^{j[2k]}Y_{j[2k-1]}{}^{\dot\alpha}
\zeta_j{}_{\dot\alpha},\nonumber
\\
\delta \bar W_{i[2k]} &=& -{2\bar\alpha_{2k}}
Y_{i[2k-1]}{}^{\dot\alpha}
\zeta_i{}_{\dot\alpha}+\frac{2\alpha_{2k}}{(2k-1)!}{\cal
E}_{i[2k]}{}^{j[2k]}Y_{j[2k-1]}{}^{\alpha} \zeta_j{}_{\alpha}.
\end{eqnarray}
In this case, again the algebra of supercharges has the form
(\ref{Q}) and the Lagrangian is invariant under the transformations
(\ref{STB}), (\ref{STF}), (\ref{STB}), (\ref{ST3/2}), (\ref{ST1}),
(\ref{ST1/2}), (\ref{ST04k}).

To conclude this subsection one notes that we have studied here only
the case of maximal integer spin $k$. In the case of maximal
half-integer spin the above consideration is not applicable since
the relation (\ref{ScalCond}) is inconsistent with the half-integer
$k$. This case requires a special analysis. The matter is that the
consistent usage of the generic scheme described the subsection
\ref{Sec4.3} leads to double of all the fields in the
supermultiplet. In the case of maximal integer spin we can avoid
this doubling if one imposes, in particular, the condition
(\ref{ScalCond}). In the case of maximal half-integer spin such a
condition is impossible\footnote{We are grateful to Yu.M. Zinoviev
for discussion of this question.}.

\section{Summary and Prospects}
In this paper we have studied the field realization of arbitrary
$N$-extended massless supermultiplets in $4D$ AdS space. For
arbitrary highest integer or half-integer spin $k$ fields entering
the supermultiplets we realized the on-shell supersymmetric
component Lagrangian formulations under the condition $N<4k$ and
defined the higher superspin field Lagrangians as the sums of free
Lagrangians for all the fields in the supermultiplets. We have
constructed the supertransformations which form the closed on-shell
algebras and leave invariant the Lagrangians. It is shown that the
commutators of two such supertransformations form the $N$-extended
AdS superalgebra, i.e. they are the combinations of the
translations, Lorentz rotations and internal
$SO(N)$-transformations. Also, we have realized the maximally
extended supermultiplets where $N=4k$ in the case of the highest
integer spin $k$ and constructed the corresponding
supertransformations.

We hope that our results can be helpful for the construction of the
extended massive higher spin and massless infinite spin
supermultiplets and their Lagrangian formulations extending the
results of the $N=1$ case \cite{BSZa,BSZb,BSZc}. Besides, we point
out some other open problems in the free supersymmetric higher spin
field theory. First of all, this is a problem of the superfield
Lagrangian formulation of $4D, N=1$ supersymmetric massive higher
superspin fields. The corresponding massless theories have been
constructed in the works \cite{KSP93,KS93,KS94}. As to massive
theories, there are only partial examples of the higher superspin
massive $N=1$ superfield models
\cite{BGLPa,BGLPb,BGLPc,BGLPd,BGLPe,BGLPf}. Problem of formulating
the extended supersymmetric higher spin theories in terms of
unconstrained superfields is completely open. At present, the only
case where such a possibility can in principle be realized is the
$4D, N=2$ supersymmetry, where harmonic superfield approach
\cite{GIOS} allows to construct the field models in terms of $N=2$
unconstrained superfields. We hope to study some of these open
problems in the forthcoming papers.

\section{Acknowledgements}
The authors thank S.M. Kuzenko for useful comments and Yu.M.
Zinoviev for fruitful discussions. The work was partially supported
by the Ministry of Science and Higher Education of the Russian
Federation, project No. FEWF-2080-0003 and RFBR grant, project No.
18-02-00153. T.V.S acknowledges partial support from the President
of Russia grant for young scientists No. MK-1649.2019.2.

\appendix

\section{Notations and conventions}

In the paper, we adopt the "condensed notation" for the indices.
Namely, if some expression contains $n$ consecutive indices, denoted
by the same letter with different numbers (e.g.
$\alpha_1,\alpha_2,\ldots \alpha_n$) and is symmetric on them, we
simply write the letter, with the number $n$ in parentheses if $n>1$
(e.g. $\alpha(n)$). For example:
\begin{equation}
\Phi^{\alpha_1,\alpha_2,\alpha_3} = \Phi^{\alpha(3)}, \qquad
\zeta^{\alpha_1} \Omega^{\alpha_2\alpha_3} = \zeta^\alpha
\Omega^{\alpha(2)}
\end{equation}
We define symmetrization over indices as the sum of the minimal
number of terms necessary without normalization multiplier.

We use the multispinor formalism in four dimensions (see e.g.
\cite{BK}). Every vector index is transformed into a pair of spinor
indices: $V^\mu\sim V^{\alpha\dot{\alpha}}$, where
$\alpha,\dot{\alpha}=1,2$. Dotted and undotted indices are
transformed into one another under the hermitian conjugation:
\begin{equation}
\left(\Omega^{\alpha{\dot{\alpha}(2)}}\right)^\dagger=\Omega^{\alpha(2){\dot{\alpha}}}
\end{equation}
The spin-tensors, i.e. fields with odd number of indices, are
Grassmannian. For example,
\begin{equation}
A^{\alpha(2)\dot\alpha} \eta^{\alpha} = - \eta^{\alpha}
A^{\alpha(2)\dot\alpha}
\end{equation}
Under the hermitian conjugation, the order of fields is reversed:
\begin{equation}
\left(A^{\alpha(2)\dot\alpha}\eta^{\alpha}\right)^\dagger =
\eta^{\alpha} A^{\alpha(2)\dot\alpha} = -
A^{\alpha(2)\dot\alpha}\eta^{\alpha}
\end{equation}
The metrics for the spinor indices is an antisymmetric bispinor
$\epsilon_{\alpha\beta}$ and inverse $\epsilon^{\alpha\beta}$:
\begin{equation}
 \epsilon_{\alpha\beta} \xi^\beta = - \xi_\alpha, \qquad
 \epsilon^{\alpha\beta} \xi_\beta = \xi^\alpha,
\end{equation}
similarly for dotted indices.

In the frame-like formalism, two bases, namely the world one and the
local one are used. We denote the local basis vectors as
$e^{\alpha\dot\alpha}$; the world indices are omitted, and all the
fields are assumed to be the differential forms. Similarly, all the
products of the forms are exterior with respect to the world
indices. In the paper, we use the basis forms, i.e. antisymmetrized
products of basis vectors $e^{\alpha\dot\alpha}$. The forms are the
2-form $E^{\alpha(2)}+h.c.$, the 3-form $E^{\alpha\dot\alpha}$ and
the 4-form $E$. The transformation law of the forms under the
hermitian conjugation is:
\begin{equation}
(e^{\alpha\dot\alpha})^\dagger=e^{\alpha\dot\alpha}, \qquad
(E^{\alpha(2)})^\dagger=E^{\dot\alpha(2)}, \qquad
(E^{\alpha\dot\alpha})^\dagger=-E^{\alpha\dot\alpha}, \qquad
(E)^\dagger=-E
\end{equation}
The covariant AdS derivative satisfies the following normalization
conditions:
\begin{equation}
D\wedge D \Omega^{\alpha(m)\dot\alpha(n)} = -
2\lambda^2(E^\alpha{}_\beta\Omega^{\alpha(m-1)\beta\dot\alpha(n)}+E^{\dot\alpha}{}_{\dot\beta}
\Omega^{\alpha(m)\dot\alpha(n-1)\dot\beta})
\end{equation}

\end{document}